\documentclass[]{proarticle}
\usepackage{multicol}
\usepackage{graphicx}
\usepackage{amsmath}
\usepackage{amssymb}
\usepackage{epsfig}

\def\nn{\nonumber}

\def\beq{\begin{equation}}
\def\eeq{\end{equation}}
\def\bea{\begin{eqnarray}}
\def\eea{\end{eqnarray}}
\def\bsub{\begin{subequations}}
\def\esub{\end{subequations}}

\begin{document}
\title{Particle number production
and time variation with non-equilibrium quantum field theory}
\author{Ryuichi Hotta $^{1}$, Hiroyuki Takata $^{2}$, Takuya  Morozumi $^{3}$}
\maketitle
\address{$^{1,3}$Graduate School of Science,
Hiroshima University, 
Higashi-Hiroshima, 739-8526, Japan \\
$^2$ Tomsk State Pedagogical University, Tomsk, 634061, Russia}
\eads{$^1$ hottarc@theo.phys.sci.hiroshima-u.ac.jp,
$^2$ takata@tspu.edu.ru,
$^3$ morozumi@hiroshima-u.ac.jp} 
\begin{abstract}
We study the particle number production and its time variation using 
non-equilibrium quantum field theory. 
We study the model proposed by Hotta et.al. \cite{Hotta:2012hi}
for particle number production 
with a heavy neutral scalar and a light complex scalar
.
The interaction Lagrangian 
contains CP violating phase and particle number violating interaction among the
scalars. The particle number violating mass term is also introduced, which 
splits a complex scalar into two real scalars with small non-degenerate mass.
Therefore, the term generates particle and anti-particle mixing. 
We study the long time behavior of the particle number production rate. 
\end{abstract}
\begin{multicols}{2}
\section{Introduction}
Study of the mechanism of the particle number production
is a very important issue in baryogenesis and leptogensis.
In experiments, there are many phenomena which 
violate the particle number. Such phenomena includes
$B \bar{B}$ mixing and neutrino flavor oscillation. 
While they can be treated with time evolution of
pure state, the baryogenesis and leptogenesis occur 
in the environment where the statistical treatment is 
suitable. This is because they occur when the definite particle number of 
the universe is unknown. In this study,
We use the non-equilibrium field theory with the
density matrix and study the time evolution of the particle number. 
This paper is organized as follows. In section II, we propose
a particle number violating model which consists of a heavy neutral 
scalar and one complex scalar. In the next section,
the current associated with the particle number is 
 written in terms of a Green function of non-equilibrium field theory.
In section III, the particle number production rate is computed 
and its property is discussed.  The final section is devoted
to conclusion.
\section{Lagrangian for the scalar model and Particle Number Production}
In the previous paper \cite{Hotta:2012hi}, we proposed the following model for
particle number production with the interaction.
\bea
{\cal L}&=&\frac{1}{2} \partial_\mu N \partial^\mu N
-\frac{M_N^2}{2} N^2 +\nn \\
&+& \partial_\mu \phi^\dagger \partial^\mu \phi-
m_{\phi}^2 \phi^\dagger 
\phi+A_\phi N \phi^\dagger \phi \nn \\
&+& B^2 \phi^2 +A  N \phi^2 + h.c.,
\eea
where $N$ is a real scalar and $\phi$ is a complex scalar.
There are two types of the interaction. One is particle number conserving
interaction which coefficient is given by $A_\phi$ and the other is 
particle number violating interaction with the coefficient $A$.
There are also two types of mass term. One of them with the coefficient $B^2$
violates the particle
number and the other one is a particle number conserving one
given by the mass term $m_\phi^2 \phi^\dagger \phi$. 
One may take a phase convention that $B^2$ is real and $A$ is 
complex.  We denote the phase A 
as $\phi_A={\rm arg.}A$ and it is a source of CP violation.
The mass term $B^2$ breaks U(1) symmetry and 
it splits one complex scalar fields
into the two mass eigenstates of real scalars.  
Introducing two real scalars as $\phi=\frac{1}{\sqrt{2}}(\phi_1+i \phi_2)$, the Lagrangian
is rewritten as,
\bea
{\cal L}&=&\frac{1}{2}(\partial_\mu N \partial^\mu N 
-M_N^2 N^2) \nn \\
&+&\frac{1}{2}\sum_{i=1}^2(\partial_\mu \phi_i \partial^\mu \phi_i 
-m_i^2 \phi_i^2)+ \sum_{ij}\phi_i {\cal A}_{ij} \phi_j N,
\eea
with $m_1^2=m_\phi^2-B^2, m_2^2=m_\phi^2+B^2$.
${\cal A}_{ij}$ is a two by two matrix and is given by,
\bea
{\cal A}=
\begin{pmatrix} |A| \cos \phi_A+\frac{A_\phi}{2} & -|A| \sin \phi_A \\
-|A| \sin \phi_A & -|A|\cos \phi_A + \frac{A_\phi}{2}  
\end{pmatrix}.
\eea
\section{Computing the current for U(1) charge; particle number}
The particle number associated with the complex field $\phi$ is 
a U(1) current.  
\bea
j_\mu&=&i 
(\phi^\dagger \partial_\mu \phi -\partial_\mu \phi^\dagger \phi)\nn \\
&=& \phi_2 \partial_\mu \phi_1-\partial_\mu \phi_2 \phi_1.
\eea
We compute the divergence 
of the U(1) current with some 
initial condition specified with the
density matrix.
\bea
\langle j_\mu(X)  \rangle ={\rm Tr}\left(j_\mu(X) \rho(0)\right),
\label{eq:U1a}
\eea
where $\rho(0)$ represents an initial quantum statistical state;
\bea
\rho(0)=\frac{\exp[-\beta (H_0-\mu L)]}{{\rm Tr}[\exp[-\beta (H_0-\mu L)]]}.
\eea
$L$ is U(1) charge given by 
$L=\int d^3 x j_0$ and $\beta$ is $\frac{1}{T}$ with 
the temperature $T$.
$\mu$ is chemical potential. We assume that at $t=0$
all the interaction terms
including U(1) breaking mass term are zero and 
$t>0$, 
suddenly they are switched on.
Therefore at t=0, Hamiltonian $H_0$ and particle number $L$
commute as;
\bea
[H_0, L]=0,
\label{eq:com}
\eea
and at later time $t>0$, the U(1) breaking terms are switched on and
the particle number is not conserved. 
\bea
[H, L] \ne 0.
\eea
In $t=0$, Eq.(\ref{eq:com}),
the density matrix can be written
by the following product,
\bea
\rho(0)=\frac{\exp(-\mu L)\exp(-\beta H_0)}{{\rm Tr}[\exp(-\mu L)\exp(-\beta H_0)]}.
\eea
With the property Eq.(\ref{eq:com}), 
one can write the density matrix even for non-zero chemical
potential case.
Using the density matrix, the initial particle number at $t=0$ 
is given as,
\bea
\langle L(0) \rangle &\equiv& {\rm Tr} L(0) \rho(0) \nn \\
                     &=& \int \frac{ V d^3 k}{(2 \pi)^3} \frac{\sinh \mu \beta}
{\cosh \beta \omega_k-\cosh \beta \mu}. 
\eea
When chemical potential is zero, the initial particle number is zero.
In the following, we use the density matrix of $\mu=0$ and focus on
the particle number production with the interaction.
%\section{Particle Number L Production }
% Environment \begin{figure} ... \end{figure} places the figure in one column.
% Environment \begin{figure*} ... \end{figure*} places the figure in two column
%\begin{figure*}[hb]\center
%\includegraphics[width=2in]{figure1.eps}
%\caption{The coordinate system.}
%\label{fig1}
%\end{figure*} 
%%%%%%%%%%%%%%%%%%%%%%%%%%%%%%%%
\section{Green function and U(1) current}
%%%%%%%%%%%%%%%%%%%%%%%%%%%%%%%%
$U(1)$ current defined in Eq.(\ref{eq:U1a}) can be written in terms of 
the Green function of non-equilibrium field theory;
\cite{Bakshi:1962dv},\cite{Bakshi:1963bn},\cite{Keldysh:1964ud},
\bea
G_{12}^{12}(x,y)&=&< \phi_1(x) \rho(0) \phi_2(y)> \nn \\
                &=&{\rm Tr} (\phi_2(y) \phi_1(x) \rho(0)). 
\eea
The lower indices distinguish the species of the light scalar
fields; $\phi_i (i=1,2)$. The upper indices distinguish whether
the operator is on the time ordered path $1$ or on anti-time ordered path $2$
in closed time path formulation \cite{Schwinger:1960qe}.  
Using the definition, the divergence of the averaged
current with the density matrix is the production rate per unit time and unit 
volume.
This can be related to the Green function
as follows,
\bea
&&\frac{\partial}{\partial X^\mu} \langle J^\mu(X) 
\rangle=(\Box_x-\Box_y) G^{12}_{12}(x,y)\Bigr{|}_{x=y=X}.\nn \\
\eea
If the space translation invariance holds, the current 
depends on only time. If this is the case, the divergence of the
current is equal to time derivative of the particle
number density,
\bea
\frac{\partial}{\partial X^\mu} \langle J^\mu(X) \rangle
=
\frac{\partial}{\partial X^0} \langle J^0(X^0) \rangle.
\eea
With the help of 2 PI (Two Particle Irreducible) formulation of the
non-equilibrium quantum field theory \cite{Calzetta:1986ey}, we derive 
Schwinger Dyson equation for
the Green functions in \cite{Hotta:2012hi}.
The Schwinger Dyson equation can be solved perturbatively
and  the divergence of U(1) current obtained in one-loop level. 
We find,
\bea
&&\frac{\partial}{\partial X^\mu} \langle J^\mu(X) 
\rangle \simeq -4|A| 
\sin \phi_A A_\phi  \nn \\
&& \int \frac{d^3 p}{2 \omega_p (2 \pi)^3} \int \frac{d^3 k}{2 \omega_k
(2 \pi)^3} \frac{1}{2 \omega_N}
\nn \\
&&\{(n_k+1) n_N (n_p+1)-n_k (n_N+1) n_p \} \nn \\
&& \{I(\omega_{2p}+\omega_{2k}-\omega_N,X^0)\nn \\
        &&- I(\omega_{1p}+\omega_{1k}-\omega_N,X^0)
\}.
\label{eq:prd}
\eea
where we have ignored the terms which is proportional to $B^2$
and $\omega_{i k}=\sqrt{m_i^2+k^2} (i=1,2)$. The time dependent
function $I$ is given as,
\bea
I(\Omega,X^0)=\frac{\cos \Omega X^0-1}{\Omega},
\eea
where $n_N$ ,$n_k$($n_p$) denote the thermal equilibrium
distribution function for N and $\phi$ given as,
\bea
n_N&=&\frac{1}{\exp(\beta \omega_N(k+p))-1}, \nn \\
n_k&=&\frac{1}{\exp(\beta \omega_k)-1}.
\eea
In the divergence of the current, first
we study the factor related to distributions,
\bea
(n_k+1) n_N (n_p+1)-n_k (n_N+1) n_p.
\eea
The first term implies the decay $N$ to two light scalars while 
the second term implies inverse decay. 
If the energy conservation in the following sense,
holds,
\bea
\omega_k+\omega_p=\omega_N(p+k),
\label{eq:econv}
\eea
then,
\bea
(n_N+1) n_k n_p =n_N(n_k+1)(n_p+1).
\eea
The particle number production is cancelled between
the decay process and inverse decay process.
Therefore the net particle number can not be produced if the
energy conservation of Eq.(\ref{eq:econv})
is satisfied.
Next we study the coupling constants in Eq.(\ref{eq:prd}). The production 
rate is proportional to CP violating phase $\phi_A$. In addition 
to CP violation, both types of the interactions, one is CP conserving 
and particle number conserving one $A_\phi$ and the other is 
CP violating and particle  
number violating one $A$ are required.
Finally we study time dependence. 
We are interested in the time length for which 
coherence of the two amplitudes in Eq.(\ref{eq:prd}) corresponding to the 
$N \rightarrow \phi_i(k) \phi_i(p) $ $(i=1,2)$ is not lost yet.  
When the coherence remains, one can use the approximation,
\bea
&&I(\omega_{2p}+\omega_{2k}-\omega_N,X^0)-
 I(\omega_{1p}+\omega_{1k}-\omega_N,X^0) \nn \\
&&\simeq \frac{\sin \Omega_0 X^0}{\Omega_0} \sin
\{(\frac{B^2}{2\omega_k}+\frac{B^2}{2 \omega_p}) X^0\},
\eea
where 
$\Omega_0=\omega_N-\omega_k-\omega_p$.
We have used the approximation, 
\bea
\omega_{2k}=\sqrt{k^2+m_\phi^2+B^2}&=&\omega_k+\frac{B^2}{2 \omega_k}, \nn \\
\omega_{1k}=\sqrt{k^2+m_\phi^2-B^2}&=&\omega_k-\frac{B^2}{2 \omega_k}. \nn
\eea
We consider when $X^0$ is large and is proportional to $\frac{1}{B^2}$.
In the limit of small $B^2$ with fixed $B^2 X^0$,
the time dependent factor becomes,
\bea
&&I(\omega_{2p}+\omega_{2k}-\omega_N,X^0)-
 I(\omega_{1p}+\omega_{1k}-\omega_N,X^0) \nn \\
&&\simeq \pi \delta(\Omega_0)
\sin\{(\frac{B^2}{2\omega_k}+\frac{B^2}{2 \omega_p}) X^0\}.
\eea
Substituting the time dependent factor, the divergence of the current 
is,
\bea
&&\frac{\partial}{\partial X^\mu} \langle J^\mu(X) 
\rangle \simeq -4|A| 
\sin \phi_A A_\phi  \nn \\
&& \int \frac{d^3 p}{2 \omega_p (2 \pi)^3} \int \frac{d^3 k}{2 \omega_k
(2 \pi)^3} \frac{1}{2 \omega_N} \pi 
\delta(\omega_N-\omega_k-\omega_p)
\nn \\
&&\{(n_k+1) n_N (n_p+1)-n_k (n_N+1) n_p \} \nn \\
&&\times  \sin\{(\frac{B^2}{2\omega_k}+\frac{B^2}{2 \omega_p}) X^0\}.
\eea
The presence of the delta function implies the energy conservation 
 $\omega_N=\omega_k+\omega_p$ holds.  Therefore the decay contribution
and inverse  
decay contribution are cancelled each other. Below,
 we consider only the
decay contribution.
\bea
&&\frac{\partial}{\partial X^\mu} \langle J^\mu(X) 
\rangle \Bigr{|}_{\rm Decay} \simeq -4 \pi |A| 
\sin \phi_A A_\phi  \nn \\
&& \int \frac{d^3 p}{2 \omega_p (2 \pi)^3} \int 
\frac{d^3 k}{2 \omega_k(2 \pi)^3} \frac{1}{2 \omega_N} 
\delta(\omega_N-\omega_k-\omega_p)
\nn \\
&&\frac{e^{\beta(\omega_k+\omega_p)}}{(e^{\beta \omega_k}-1)
(e^{\beta \omega_k}-1)(e^{\beta \omega_N}-1)}
\nn \\
&& \times \sin\{(\frac{B^2}{2\omega_k}+\frac{B^2}{2 \omega_p}) X^0\}
\nn \\
&&\simeq -\frac{1}{64 \pi^3} |A| A_\phi T^2 \sin \phi_A \nn \\
&& \int_{\beta m_N}^\infty \frac{dU}{\sinh \frac{U}{2}}
\int_0^{V_{max}(U)} dV 
\frac{\sin \frac{2 \beta B^2 X^0 U}{U^2-V^2}}{\cosh\frac{U}{2}-
\cosh \frac{V}{2}}, \nn \\ 
\eea
where
\bea
U=\beta(\omega_k+\omega_p), \quad V=\beta(\omega_k-\omega_p),
\eea
and
\bea
V_{max}=\sqrt{(1-\frac{4 m_\phi^2}{m_N^2})(U^2-\beta^2m_N^2)}.
\eea
We introduce the following dimensionless quantities,
\bea
\hat{\beta}=\beta m_{\phi},
\hat{X^0}=m_\phi X^0,
\hat{B}=\frac{B}{m_\phi}, \hat{m}_N=\frac{m_N}{m_\phi}.
\label{eq:dimless}
\eea
To study the long time behaviour of the production rate
$\hat{X}^0 \sim \frac{1}{\hat{B}^2}$, 
we define the rescaled time $t$,
\bea
t=\frac{B^2 X^0}{m_{\phi} \pi} =\frac{\hat{B}^2 \hat{X}^0}{\pi}.
\label{eq:rescaledt}
\eea
Using Eq.(\ref{eq:rescaledt}), one may write 
the time dependent part as,
\bea
F[t, \hat{\beta}]&=&\frac{1}{\hat{\beta}^2}
\int_{\hat{\beta}\hat{m}_N}^\infty \frac{dU}{\sinh \frac{U}{2}} \nn \\
&\times &\int_0^{V_{max}(U)} dV 
\frac{\sin \frac{2 \pi \hat{\beta} U t}{U^2-V^2}}{\cosh\frac{U}{2}-\cosh\frac{V}{2}}.
\label{eq:td}
\eea
In 
Fig.1, we show the time dependent factor Eq.(\ref{eq:td}) of the
production rate as a function of the
dimensionless rescaled time $ t$ for three inverse temperature
$\hat{\beta}=\frac{1}{10},\frac{1}{20}$,and $\frac{1}{30}$.
We choose $\hat{m}_N=20$. 
We can see the the production rate oscillates and the amplitude decreases.
As temperature grows, the production rate becomes larger. 
\begin{figure*}[th]\center
\includegraphics[width=7cm]{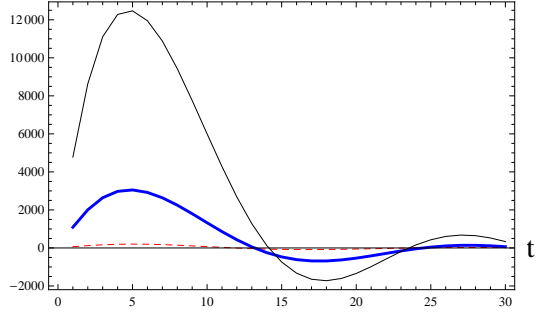}
\caption{Time dependence of the production rate
of the particle number density. The dashed line 
corresponds to $\hat{\beta}=\frac{1}{10}$, the thick
solid line corresponnds to $\hat{\beta}=\frac{1}{20}$
, and the thin solid line corresponds to 
$\hat{\beta}=\frac{1}{30}$.}
\end{figure*}
%%%%%%%%%%%%%%%%%%%%%%%%%%%%%%%%
\section{Conclusion}
\begin{itemize}
\item{We compute the time variation of the particle number with a scalar model.}\item{In the model, a neutral scalar and one complex scalar are included.
U(1) charge related to the complex scalar is the particle number.}
\item{The particle number and CP symmetry 
are violated due to the mass term and the interaction.}
\item{Due to U(1) soft-breaking term, one complex scalar is not a mass
eigenstate and mass eigenstates are two real scalars with non-degenerate 
mass.} 
\item{We have computed the time dependence of the production rate
of the particle number density.}
\item{The rate oscillates and is damping in their amplitude. The oscillation
period is larger for the smaller non-degeneracy.} 
\item{Because of the damping, the particle number produced during the first
half cycle of the oscillation,
may remain after the integration of the rate with respect to time.} 
\item{We have not included the effect of the expanding universe,
which effect leads to the decay and inverse decay contribution
may not cancel.}
\item{As an extension of our work, one can consider the case for $<L(0)>\ne0$
and how the initial particle number is washed out. Also we may apply the method
to flavored leptogenesis; $L_e(t), L_\mu(t), L_\tau(t)$.}
\end{itemize}
%%%%%%%%%%%%%%%%%%%%%%%%%%%%%%%%
\section*{Acknowledgement}
This work was supported by Grant-in-Aid for Scientific Research (C)
,Grant Number 22540283 from JSPS. We thank organizers and participants of 
QFTG2012.  
%%%%%%%%%%%%%%%%%%%%%%%%%%%%%%%%

\end{multicols}
\end{document}